\documentclass{article}

\usepackage{times}
\usepackage{epsfig}
\usepackage{graphicx}
\usepackage{amsmath}
\usepackage{amssymb}
\usepackage{multirow}
\usepackage{paralist}
\usepackage[table]{xcolor}
\usepackage{subfigure}

\usepackage[pagebackref=true,breaklinks=true,colorlinks,bookmarks=false]{hyperref}

\begin{document}

\title{TEyeD: Over 20 million real-world eye images with Pupil, Eyelid, and Iris 2D and 3D Segmentations, 2D and 3D Landmarks, 3D Eyeball, Gaze Vector, and Eye Movement Types}

\author{Wolfgang fuhl\\
University Tübingen\\
{\tt\small Wolfgang.Fuhl@uni-tuebingen.de}
\and
Gjergji Kasneci\\
Technischen Universität München\\
{\tt\small Gjergji.Kasneci@tum.de}
\and
Enkelejda Kasneci\\
Technischen Universität München\\
{\tt\small Enkelejda.Kasneci@tum.de}
}

\maketitle

\begin{abstract}
   We present TüEyeD, the world's largest unified public data set of eye images taken with head-mounted devices. TüEyeD was acquired with seven different head-mounted eye trackers. Among them, two eye trackers were integrated into virtual reality (VR) or augmented reality (AR) devices. The images in TüEyeD were obtained from various tasks, including car rides, simulator rides, outdoor sports activities, and daily indoor activities. The data set includes 2D\&3D landmarks, semantic segmentation, 3D eyeball annotation and the gaze vector and eye movement types for all images. Landmarks and semantic segmentation are provided for the pupil, iris and eyelids. Video lengths vary from a few minutes to several hours. With more than 20 million carefully annotated images, TüEyeD provides a unique, coherent resource and a valuable foundation for advancing research in the field of computer vision, eye tracking and gaze estimation in modern VR and AR applications.\\
Download: \url{https://es-cloud.cs.uni-tuebingen.de/d/8e2ab8c3fdd444e1a135/?p=%2FTEyeDS&mode=list}.\\
Alternative Download: \url{https://hctlsrva.edu.sot.tum.de/TEyeDS/}.\\
\end{abstract}

\section{Introduction}

Image-based eye tracking is becoming increasingly important in today's world, as human eye movements~\cite{FCDGR2020FUHL,fuhl2018simarxiv,ICMIW2019FuhlW1,ICMIW2019FuhlW2,EPIC2018FuhlW} and scan path~\cite{C2019,FFAO2019} have the potential to revolutionize the way we interact with computer systems around us~\cite{hutchinson1989human,UMUAI2020FUHL} and give insights into the visual process of humans~\cite{AGAS2018,DWTE022017}. Since our actions and intentions can be recognized and - to a certain degree - anticipated from the way we move our eyes, eye movement analysis can enable completely new applications, especially when coupled with modern display technologies like VR or AR. For example, the gaze signal, together with the associated possibility of human-machine interaction, enables people with disabilities to interact with their environment through the use of special devices tailored to the patient's disability~\cite{adjouadi2004remote}. In the case of surgical microscopes where the surgeon has to operate a multitude of controls, the visual signal can be used for automatic focusing~\cite{CORR2017FuhlW1,WTCDAHKSE122016,032017,0320170,Bahmani2016}. Furthermore, in scenarios where it is important to identify the expertise of a person (e.g., surgery, image interpretation, etc.) the gaze signal can be used together with interaction patterns to predict the expertise of a subject in a given task~\cite{032017,ACTNEURO2017}. Gaze behavior can also be used to diagnose a variety of diseases~\cite{sturm2011mutual}, such as  schizophrenia~\cite{hooker2005you}, autism~\cite{dalton2005gaze}, Alzheimer's disease ~\cite{sturm2011mutual}, glaucoma~\cite{ishiyama2015usefulness}, and many more. Additionally, in VR/AR and gaming, the gaze signal can be used to reduce the computations of rendering resources~\cite{patney2016towards}. 

\textbf{A look at the human eye beyond the gaze information} opens up further sources of information. For example,  The gaze signal alone, however, is by no means the limit to information offered by the human eye~\cite{WF042019}. The 
the frequency of eyelid closure, can be used to measure a person's fatigue~\cite{dong2005fatigue}, an effective safety feature in driving~\cite{dong2005fatigue} and aviation~\cite{dinges2005pilot} scenarios. Of course, this applies to all safety critical tasks that are monitored by one or more persons~\cite{luckiesh1937eyelid}. Another significant source of information is the pupil size, which can serve as a basis to estimate the cognitive load of a person in a given task~\cite{duchowski2018index}. This may then be used for better adaption of the content (e.g. in media-based learning) to a person's mental state. 
Finally, eye-related information can be used in identification processes~\cite{C2019,FFAO2019}, not only through the unique imprint of the iris~\cite{boles1998security}, but also through an individual's gaze behavior~\cite{C2019,FFAO2019}. 

In the age of machine learning~\cite{RLDIFFPRIV2020FUHL,AAAIFuhlW, NNPOOL2020FUHL, RINGRAD2020FUHL}, where there is an abundance of effective and scalable learning approaches ~\cite{lecun2015deep,rosenblatt1958perceptron,scholkopf2002learning,breiman2001random,chen2016xgboost, friedman2002stochastic, freund1996experiments}, it is, in principle, easier to develop algorithms or models which automatically retrieve the necessary information directly from the data. However, carefully annotated and curated data remains the central prerequisite for the development of machine learning - and especially deep learning - methods. 

Providing such a prerequisite for a broad range of scenarios that involve eye-related information is exactly what the TüEyeD data set aims to achieve. 

\textbf{Our contribution to the state-of-the-art} is as follows:
\begin{compactitem}
	\item[1] We provide the largest, unified, richly and coherently labeled data set of over 20 million eye images, collected using seven different eye trackers, ranging from 25 Hz to 200 Hz, including integration in VR- and AR-scenarios. 
	\item[2] Moreover, TüEyeD covers a wide range of tasks and activities, such as car driving, driving in a simulator, indoor and outdoor activities including VR and AR scenarios,
	\item[3] and provides careful annotations for 2D\&3D segmentation, 2D\&3D landmarks, pupil center, the gaze vector, the eyeball model, and eye movement types (fixations or saccades).
	\item[4] TüEyeD was generated from recordings in real-world settings, thus containing a wide range of realistic challenges, such as different resolutions, steep viewing angles, varying lighting conditions or device slippage.\\
\end{compactitem}

\textbf{Broader impact of TüEyeD.} We believe that our data set will boost the research and the application of gaze-related information in various fields. In \textit{computer vision and machine learning}, TüEyeD can serve as an annotated data set for the development of (deep learning) models on eye segmentation, iris recognition and feature extraction. In \textit{eye-tracking research}, TüEyeD can help to improve the quality of eye position and gaze estimation models and the automated classification of eye movement types.
In \textit{cognitive sciences}, TüEyeD can serve as a foundation for the development of models that assess the user's cognitive state, such as cognitive load or fatigue, based on observable measures, such as pupil fluctuations, eye closure, eye landmarks or eye movements. In the areas of \textit{VR and AR}, eye tracking is considered a key enabling technology for two reasons. First, reliable gaze prediction can save substantial computational and rendering resources. Second, and most importantly, inferring the user's intention can enable intuitive and immersive user experiences which adapt to the user requirements during interaction. Such application,s however, require as stated by Cavin at al.\footnote{\url{https://research.fb.com/wp-content/uploads/2019/05/Eye\_Tracking\_VR\_AR\_Proposal.pdf}} \textit{".. eye-tracking to reliably work all the time for all individuals under all
environmental conditions within the power and computational constraints imposed by the form factor of VR and AR devices"}.


\section{Related work}
Various eye-image data sets generated by eye trackers already exist~\cite{kim2019nvgaze,garbin2020dataset,WTCKWE092015,WTTE032016,CORR2017FuhlW2,CORR2016FuhlW,swirski2012robust,tonsen2016labelled} including recordings from driver studies as well as simulators~\cite{WTCKWE092015,WTTE032016,CORR2017FuhlW2,CORR2016FuhlW}. In addition, there are also recordings from specific challenges ~\cite{tonsen2016labelled,swirski2012robust}. While early data sets provided only the annotated pupil center~\cite{kim2019nvgaze,garbin2020dataset,WTCKWE092015,WTTE032016,CORR2017FuhlW2,CORR2016FuhlW,swirski2012robust,tonsen2016labelled}, newer data sets offer the segmented pupil, iris, and sclera~\cite{garbin2020dataset,ICCVW2019FuhlW,CAIP2019FuhlW}, eventually extended by the optical vector~\cite{NNETRA2020} which allows for a shift in invariant gaze estimation~\cite{swirski2013fully}. Such data sets are available for conventional eye trackers~\cite{kim2019nvgaze,garbin2020dataset,WTCKWE092015,WTTE032016,CORR2017FuhlW2,CORR2016FuhlW,swirski2012robust,tonsen2016labelled} and for VR~\cite{kim2019nvgaze,garbin2020dataset} and AR~\cite{kim2019nvgaze}. In addition to these annotations, segmented iris data sets with subject identification numbers are available for the development of personal identification systems~\cite{phillips2007comments,proenca2009ubiris,proencca2005ubiris}. Other annotations that contain important information are eye movement types such as fixation, saccades, and smooth pursuits~\cite{kothari2020gaze}. However, in contrast to TüEyeD, all the mentioned data sets have a narrow, task-specific focus. 

Since the manual annotation of eye images and eye movements is very complex, especially when a high accuracy is required, several procedures to generate synthetic data have been proposed~\cite{kim2019nvgaze,swirski2014rendering,wood2015rendering}. This includes synthesized image data~\cite{kim2019nvgaze}, automated rendering methods~\cite{swirski2014rendering,wood2015rendering}, and eye movement simulations~\cite{fuhl2018simarxiv,EPIC2018FuhlW,FCDGR2020FUHL}, as well as generative adversarial networks (GAN)~\cite{ICCVW2019FuhlW}. The disadvantage of synthetic data sets is that they cannot represent relevant challenges of real-world imaging, e.g., with regard to varying illumination conditions, physiological properties of the eyes, lighting sources, device slippage, and more. This is still an important part of research today, particularly in the development of novel interaction techniques and applications in VR/AR. 

In summary, current published data sets are limited due to their focus on specific problems. To the best of our knowledge, there is no unified and coherent data set containing all the relevant annotations on eye-related information. Moreover, all data sets are generated using one specific type of eye-tracking device. In contrast, TüEyeD offers a carefully, coherently and richly labeled data set containing all relevant eye-related information on a wide range of tasks (such as car driving, driving in a simulator, indoor and outdoor activities including VR and AR scenarios). The tasks, in total, were recorded by seven different eye trackers with different recording frequencies (i.e. sampling rates ranging from 25 to 200 Hz): One for VR, one for AR, and five more from head-mounted devices. In addition, TüEyeD was generated from recordings containing a wide range of realistic challenges, such as different resolutions, steep viewing angles and varying lighting conditions or device slippage during outdoor and sports activities, to name a few. Theses challenges are known to be the limiting factor of eye-tracking in many real-world applications~\cite{fuhl2016pupil}.

\section{Comparison with Existing Data Sets}

\begin{table*}[t]
	\begin{footnotesize}
		\begin{center}
			\caption{A list of the published data sets for virtual reality (VR), augmented reality (AR), and head mounted (HM) eye tracker. The table contains information on the number of subjects (Sub.), the type of eye tracker (AR,VR,HM), the acquisition frequency (FRQ), image resolution (Res.), the number of annotated images (Num. Annot), whether or not segmentations are present in 2D\&3D (Seg 2D, Seg 3D), whether or not the pupil center is annotated (PC), whether or not landmarks are present in 2D\&3D (LM 2D,LM 3D), whether the position and radius of the eyeball is given (Eye), whether  or not the gaze vector or gaze position is given (Ga), and whether or not the eye movement types are annotated (Mov.). The subtypes stand for $I=Iris$, $P=Pupil$, $Sc=Sclera$, $Lid=Eyelid$, $F=Fixation$, $S=Saccade$, $SP=Smooth~Pursuits$, and $B=Blinks$.}
			\label{tbl:newmulti}
			\setlength\tabcolsep{1pt}
			\begin{tabular}{c|c|ccc|c|c|c|ccc|ccc|c|ccc|ccc|c|c|cccc}
				\\\hline
				Data & Sub. & \multicolumn{3}{|c|}{Tracker} & FRQ & Res. & Num. & \multicolumn{3}{|c|}{Seg 2D} & \multicolumn{3}{|c|}{Seg 3D}  & PC & \multicolumn{3}{|c|}{LM 2D} & \multicolumn{3}{|c|}{LM 3D} & Eye & Ga & \multicolumn{4}{|c}{Mov.} \\
				& & VR & AR & HM &  &  & Annot & P & I & Sc & P & I & Sc &  & P & I & Lid & P & I & Lid &  &  & F & S & SP & B \\ \hline
				POG\cite{mcmurrough2012eye} & 20 & - & - & 1 & 30Hz & $768\times480$ & - & - & - & -& - & - & - & - & - & - & - & - & - & - & - & Y & - & - & - & - \\
				NNVEC\cite{NNETRA2020} & 20 & - & - & 1 & 25Hz & $384\times288$ & 866,069 & - & - & -& - & - & - & - & - & - & - & - & - & - & Y & Y & - & - & - & - \\
				NVGaze\cite{kim2019nvgaze} & 35 & 1 & 1 & - & 120Hz & $640\times480$ & 2,500,000  & - & - & -& - & - & - & - & - & - & - & - & - & - & Y & Y & - & - & - & - \\
				Casia.v1\cite{phillips2007comments,tan2010efficient} & 108 & - & - & 1 & - & $320\times280$ & 756  & - & Y & -& - & - & - & - & - & - & - & - & - & - & - & - & - & - & - & - \\
				Casia.v2\cite{phillips2007comments,tan2010efficient} & 60 & - & - & 2 & - & $640\times480$ & 2,400  & - & Y & -& - & - & - & - & - & - & - & - & - & - & - & - & - & - & - & - \\
				Casia.v3\cite{phillips2007comments,tan2010efficient} & $\approx$700 & - & - & 3 & - & Multiple & 22,034  & - & Y & -& - & - & - & - & - & - & - & - & - & - & - & - & - & - & - & - \\
				Casia.v4\cite{phillips2007comments,tan2010efficient} & $\approx$1,800 & - & - & 4 & - & Multiple & 54,601  & - & Y & -& - & - & - & - & - & - & - & - & - & - & - & - & - & - & - & - \\
				Casia.test\cite{phillips2007comments,tan2010efficient} & 1,000 & - & - & 1 & - & $640\times480$ & 10,000 & - & Y & -& - & - & - & - & - & - & - & - & - & - & - & - & - & - & - & - \\
				Casia.age\cite{wild2015impact,bergmuller2014impact} & 50 & - & - & 2 & - & Multiple & $\approx$160,000  & - & Y & -& - & - & - & - & - & - & - & - & - & - & - & - & - & - & - & - \\
				Ubiris.v1\cite{proencca2005ubiris} & 241 & - & - & Y & 30Hz & Multiple & 877  & - & Y & - & - & - & - & - & - & - & - & - & - & - & - & - & - & - & - & - \\
				Ubiris.v2\cite{proenca2009ubiris} & 261 & - & - & Y & 200Hz & $400\times300$ & $\approx$11,000  & - & Y & - & - & - & - & - & - & - & - & - & - & - & - & - & - & - & - & - \\	
				MASD\cite{das2017sserbc,das2016ssrbc,das2019sclera} & 82 & - & - & Y & - & Multiple & 2,624  & - & - & Y& - & - & - & - & - & - & - & - & - & - & - & - & - & - & - & - \\
				GAN\cite{ICCVW2019FuhlW} & 22 & - & - & 1 & 120Hz & $640\times480$ & 130,856 & Y & - & Y  & - & - & - & - & - & - & - & - & - & - & - & - & - & - & - & - \\
				500k\cite{CAIP2019FuhlW} & 20 & - & - & 1 & 25Hz & $384\times288$ & 866,069 & Y & - & Y  & - & - & -& - & - & - & - & - & - & - & - & - & - & - & - & - \\
				OpenEDS\cite{garbin2020dataset} & 152 & 1 & - & - & 200Hz & $400\times640$ & 356,649 & Y & Y & Y  & - & - & -& - & - & - & - & - & - & - & - & - & - & - & - & - \\
				GIW\cite{kothari2020gaze} & 19 & - & - & 1 & 120Hz & $640\times480$ & $\approx$2,016,000  & - & - & -& - & - & - & - & - & - & - & - & - & - & - & - & Y & Y & Y & Y \\
				BAY\cite{santini2016bayesian} & 6 & - & - & 1 & 30Hz & $640\times480$ & 27,022  & - & - & -& - & - & - & - & - & - & - & - & - & - & - & - & Y & Y & Y & - \\
				HEV\cite{david2018dataset} & 57 & 1 & - & - & 24-30Hz & - & no images & - & - & -& - & - & - & - & - & - & - & - & - & - & - & - & Y & Y & - & - \\
				HEI\cite{rai2017dataset} & 63 & 1 & - & - & 60Hz & - & no images & - & - & -& - & - & - & - & - & - & - & - & - & - & - & - & Y & Y & - & - \\
				LPW\cite{tonsen2016labelled} & 22 & - & - & 1 & 120Hz & $640\times480$ & 130,856  & - & - & -& - & - & - & Y & - & - & - & - & - & - & - & - & - & - & - & - \\
				Swi\cite{swirski2012robust} & 2 & - & - & 1 & - & $620\times460$ & 600  & - & - & -& - & - & - & Y & - & - & - & - & - & - & - & - & - & - & - & - \\
				ExCuSe\cite{WTCKWE092015} & 7 & - & - & 1 & 25Hz & $384\times288$ & 39,001  & - & - & -& - & - & - & Y & - & - & - & - & - & - & - & - & - & - & - & - \\
				Else\cite{WTTE032016} & 17 & - & - & 1 & 25Hz & $384\times288$ & 55,712 & - & - & - & - & - & - & Y & - & - & - & - & - & - & - & - & - & - & - & - \\
				PNET\cite{CORR2017FuhlW2,CORR2016FuhlW} & 5 & - & - & 1 & 25Hz & $384\times288$ & 41,217  & - & - & -& - & - & - & Y & - & - & - & - & - & - & - & - & - & - & - & - \\
				EWO\cite{WTDTWE092016} & 11 & - & - & 1 & 25Hz & $384\times288$ & 1,100   & - & - & - & - & - & - & - & - & - & Y & - & - & - & - & - & - & - & - & - \\
				FRE\cite{WTE032017} & 11 & - & - & 1 & 25Hz & $384\times288$ & 4,000 & - & - & - & - & - & - & - & - & - & Y & - & - & - & - & - & - & - & - & - \\ \hline
				\multirow{4}{*}{\rotatebox{90}{TüEyeD}} & 39 & - & - & 1 & 25Hz & $384\times288$ & 5,665,053 & Y & Y & Y & Y & Y & Y & Y & Y & Y & Y & Y & Y & Y & Y & Y & Y & Y & Y & Y \\ 
				& 1 & - & - & 1 & 60Hz & $320\times240$ & 12,184 & Y & Y & Y & Y & Y & Y & Y & Y & Y & Y & Y & Y & Y & Y & Y & Y & Y & Y & Y\\
				& 22 & - & - & 1 & 95Hz & $640\times480$ & 130,856 & Y & Y & Y & Y & Y & Y & Y & Y & Y & Y & Y & Y & Y & Y & Y & Y & Y & Y & Y\\
				& 54 & 1 & 1 & 1 & 120Hz & $640\times480$ & 8,691,764 & Y & Y & Y & Y & Y & Y & Y & Y & Y & Y & Y & Y & Y & Y & Y & Y & Y & Y & Y\\
				& 16 & - & - & 1 & 60Hz & $640\times360$ & 6,367,216 & Y & Y & Y & Y & Y & Y & Y & Y & Y & Y & Y & Y & Y & Y & Y & Y & Y & Y & Y\\ \hline
			\end{tabular}
		\end{center}
	\end{footnotesize}
\end{table*}

\begin{figure*}[h]
	\centering
	\includegraphics[width=0.72\textwidth]{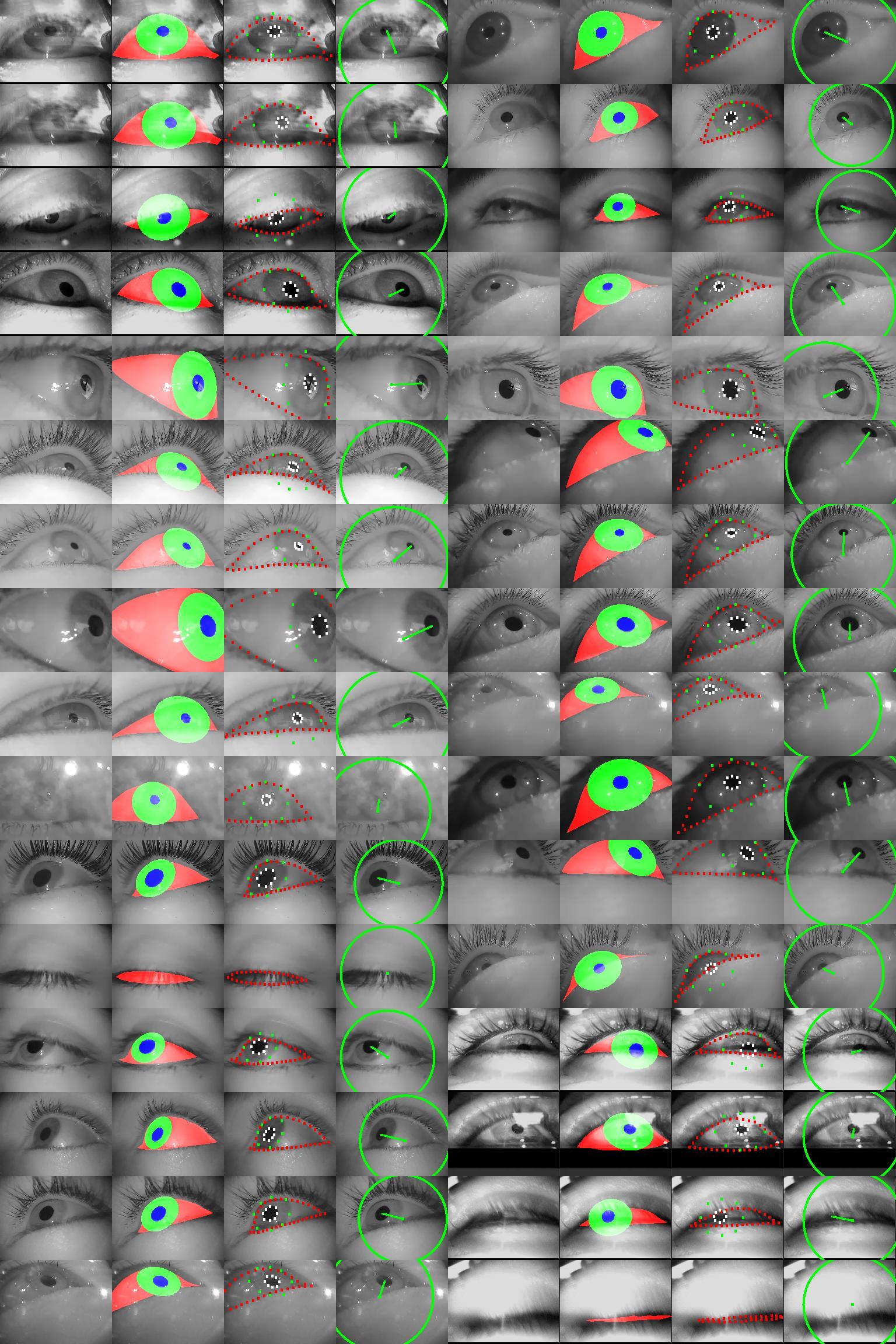}
	\caption{Example images from our data set with annotations.}
	\label{fig:imagesds}
\end{figure*}

Table~\ref{tbl:newmulti} provides an overview of existing data sets containing close-up eye images. Each data set deals with a specific issue, such as Casia and Ubiris, which are used to identify individuals by the iris. Direct gaze estimation, as in POG, NVGaze or NNVEC, is tackled by a more recent group of data sets. In NNVEC, the direct estimation of the optical vector and eyeball position make it possible to compensate for shifts of the head mounted eye tracker. In contrast to Casia and Ubiris, MASD focuses on the segmentation of the eye's sclera. MASD can be used to improve iris segmentation while also helping to estimate the degree of eye opening, an indicator of blink rate. Different eye movement types are offered alongside images in GIW and BAY while additional annotations of eye movement types for worn eye trackers are published in HEV and HEI. Proving to be very challenging, GAN and 550k existing data sets, LPW, ExCuSe, Else, and PNET, were extended with segmentations for the pupil~\cite{062016,WDTE092016,WTCDAHKSE122016,WTCDOWE052017,WDTTWE062018,VECETRA2020,CORR2017FuhlW1,ETRA2018FuhlW,ICCVW2019FuhlW,CAIP2019FuhlW,ICCVW2018FuhlW} and sclera~\cite{WTDTWE092016,WTDTE022017,WTE032017}. Originally, these data sets, together with the Swi data set, only provided the pupil center as annotation. OpenEDS was the first data set with segmentations for the pupil, iris, and sclera. Containing many subjects, OpenEDS was specifically acquired to enable VR-related research and applications. Additionally, two data sets (EWO and FRE) encompass 2D landmarks specific to worn eye trackers and were published together with real-time algorithms for the CPU. 
TüEyeD both combines and extends previously published data sets by utilizing seven different eye trackers, each with a different resolution, incorporating all available annotations offered by existing data sets, and broadening these sets with 3D segmentation and landmarks. More specifically, the data sets integrated in TüEyeD are NNGaze, LPW, GIW, ElSe, ExCuSe, and PNET. Additionally, the complete data from the study~\cite{kasneci2014driving} was also carefully annotated. Additional annotated data was recorded with an eye tracker from Enke GmbH and the eye tracker from Look!. In total, TüEyeD contains more than 20 million images, making it, to our knowledge, the world's largest data set of images taken by head mounted eye trackers. Similar to the OpenEDS data set, some data (6,379,400 samples collected from the Look! and Enke GmbH eye trackers) was withheld from TüEyeD in order to obtain a reliable evaluation beyond generalization of single eye trackers. By deploying trained models or pre-compiled programs, it is possible to achieve fair evaluation of the runtime and generalization of different eye trackers. Unfortunately, we could not consider eye images captured with Tobii eye trackers due to prohibitory license agreements.

\section{Further Data Set Details}
\begin{figure*}[t]
     \centering
     \begin{subfigure}
         \centering
         \includegraphics[width=0.3\textwidth]{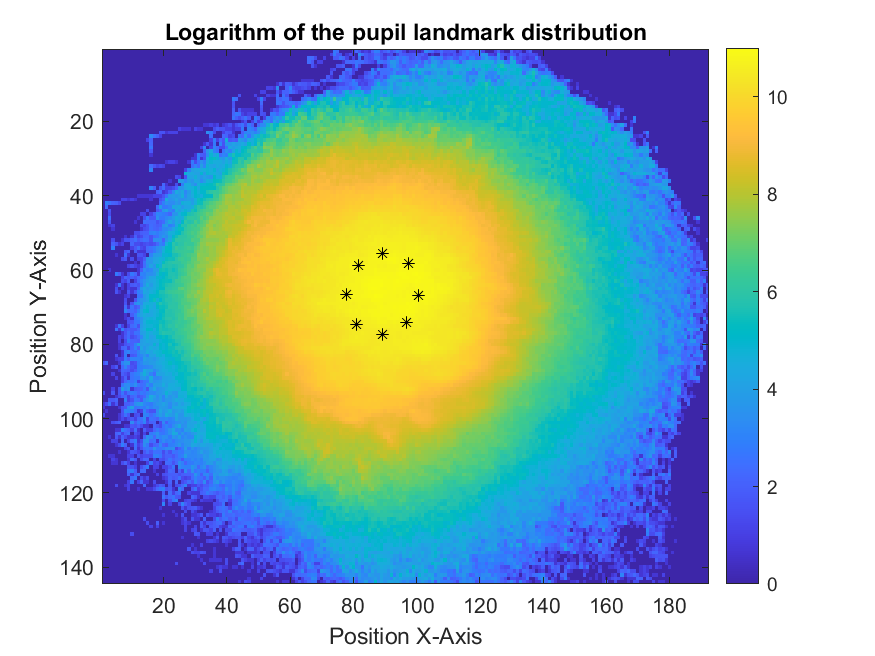}
         \label{fig:lmpupil}
     \end{subfigure}
     \begin{subfigure}
         \centering
         \includegraphics[width=0.3\textwidth]{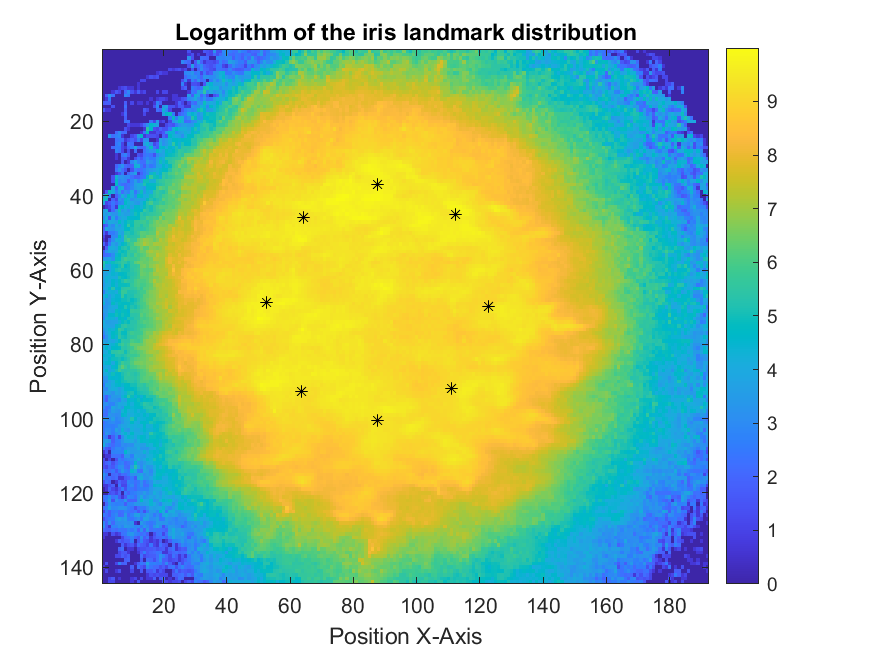}
         \label{fig:lmiris}
     \end{subfigure}
     \begin{subfigure}
         \centering
         \includegraphics[width=0.3\textwidth]{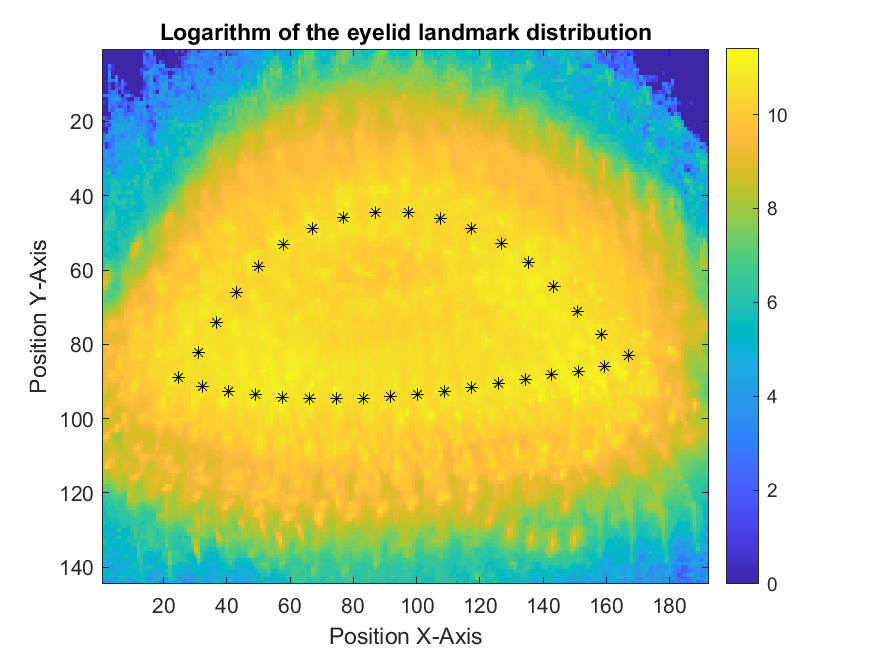}
         \label{fig:lmeyelid}
     \end{subfigure}
    \caption{The logarithmic distribution of the pupil landmarks (left), iris landmarks (middle), and eyelid landmarks (right) in TüEyeD.}
    \label{fig:lmpupil-eyelid-iris}
\end{figure*}
Figure~\ref{fig:imagesds} shows sample images of TüEyeD. The first and fifth column contain the input images. The second and sixth column show these images with overlaid segmentations of the sclera, iris and pupil. The third and seventh column show the landmarks on the input image with red landmarks belonging to the eyelids, green landmarks to the iris, and white landmarks to the pupil. In the fourth and eighth column, the calculated eyeball is displayed as well as the center of the eyeball and the gaze vector.

Table~\ref{tbl:generalstats} shows rough statistics from the TüEyeD data set. 
Interestingly, our data set contains also images in which no eye is present. This can occur when the eye tracker is removed from the subject or when reflections in the near-infrared rage of the subject's glasses are so strong that the eye is no longer visible. Finally, our data set contains images in which the pupil is annotated but no iris appears. Such images can occur when the eye tracker is removed from the subject or when reflections in the near-infrared rage of the subject's glasses are so strong that the eye is no longer visible.
\begin{table}[h]
\begin{small}
	\begin{center}
		\caption{General statistics of our data set.}
		\label{tbl:generalstats}
		\begin{tabular}{rc}
		\\ \hline
			Pupils & 19,927,927 \\
			Iris & 19,756,546 \\
			Eyelid & 20,666,096 \\
			No eye images & 200,977 \\
			Open eyes & 19,859,456\\
			Closed eyes & 806,640 \\ 
			Blink images & 1,890,816 \\\hline
		\end{tabular}
	\end{center}
	\end{small}
\end{table}




Figures~\ref{fig:lmpupil-eyelid-iris} 
 shows the logarithmic distribution of landmarks for the pupil (left), iris (middle), and eyelids (right). Since a large part of our image data comes from real images, we used the logarithm to show all occurrences, since this enables a better representation of the areas, which are underrepresented in normal gaze behavior. One such occurrence, for example, could be a subject driving a car. The logarithm accounts for the driver's view mainly being directed forward. Hereby, our data set can also be used to evaluate tracking algorithms. In addition to the logarithmic distribution, black crosses represent the mean position of all landmarks in Figures~\ref{fig:lmpupil-eyelid-iris}, 
which are distributed over almost the entire image area. There are also individual landmarks located outside of the image area, especially in the corners of the eyelids and in the upper landmarks of the iris, as can be seen in Figure~~\ref{fig:lmpupil-eyelid-iris}.

\begin{figure}[h!]
	\centering
	\includegraphics[width=0.38\textwidth]{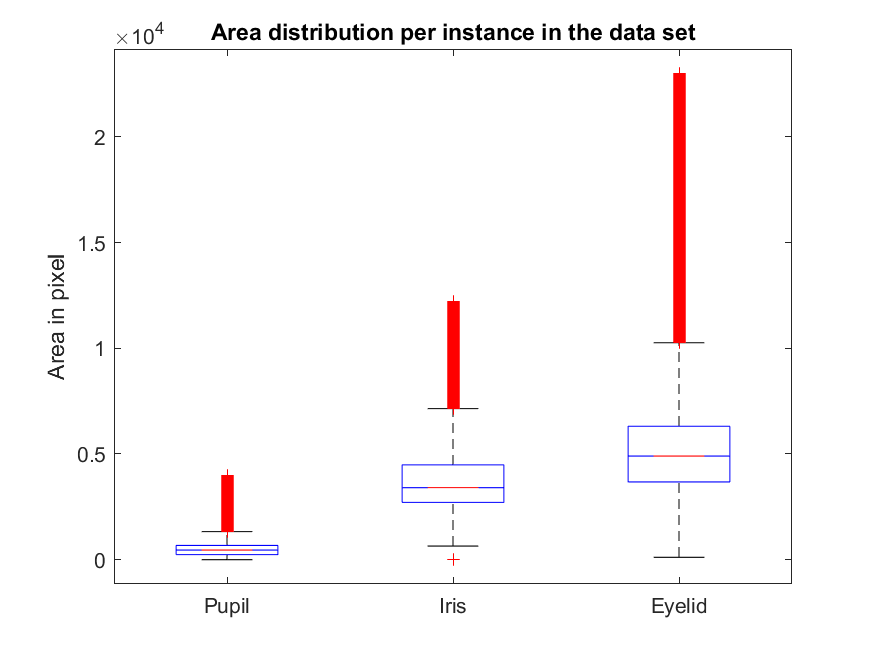}
	\caption{The area distribution for the pupil, iris, and eyelids on an $192\times144$ image resolution. The blue box corresponds to the 25th and 75th percentiles. Red crosses are the outlier and the red line corresponds to the median.}
	\label{fig:areadistri}
\end{figure}

Figure~\ref{fig:areadistri} shows the area distribution (in pixels) of the pupil, iris, and the sclera as a whisker plot. The blue boxes represent the confidence intervals with the 25th and 75th percentiles. In the middle of the blue box, the red line represents the median. The red crosses represent. As exhibited here, our data set contains different camera distances due to the specificities of the individual eye trackers and the LPW data set's special recordings which were taken at close range. TüEyeD also incorporates both large and small pupils, the result of different camera distances as well as a variety of lighting conditions.


\begin{figure}[h]
	\centering
	\includegraphics[width=0.33\textwidth]{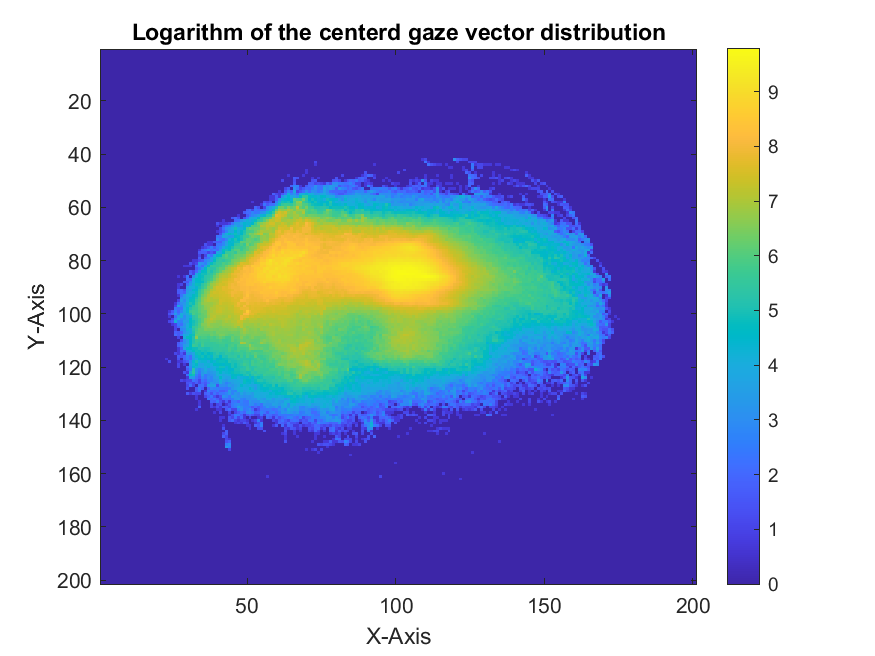}
	\caption{The logarithmic distribution of the gaze vector centered and mapped to a unit sphere in our data set.}
	\label{fig:gazedistri}
\end{figure}
Figure~\ref{fig:gazedistri} shows the logarithmic gaze vector distribution, where all vectors are unit vectors and shifted to the same center. As this figure incorporates close-up images of the eye, the depth of every vector favors the direction of the camera. Thus, depth information is not shown separately. We decided to use a logarithmic representation, similar to the landmarks, because the gaze is typically consistent and centrally aligned during activities such as driving a car. This also allows for the evaluation of tracking algorithms on TüEyeD. As shown in Figure~\ref{fig:gazedistri}, the gaze vector is distributed over the entirety of the eye ball hemisphere.

\begin{figure}[h]
	\centering
	\includegraphics[width=0.38\textwidth]{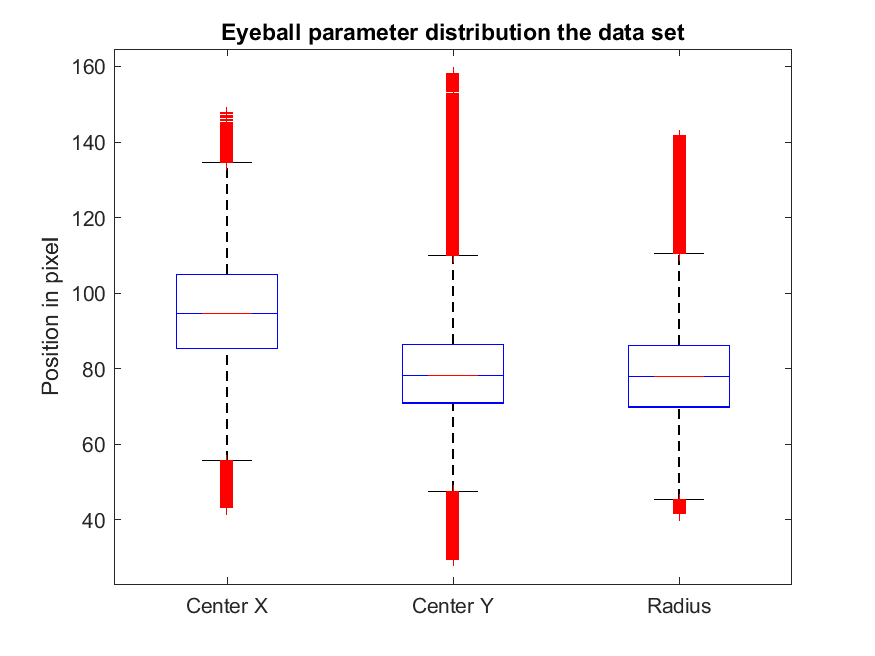}
	\caption{The distributions for the eyeball center x and y as well as the distribution of the eyeball radius in our data set on a $192\times144$ image resolution. The blue box corresponds to the 25th and 75th percentiles. Red crosses are the outlier and the red line corresponds to the median.}
	\label{fig:eyeballdistri}
\end{figure}

Figure~\ref{fig:eyeballdistri} shows the eyeball position (x,y) distribution as well as the eyeball radius in pixels as a whisker plot mapped to a fixed resolution of $192\times144$. 
The z position is not shown because we set the z position for the eye balls to zero. This means that the eyeball center is the origin for the 3D positions of the landmarks and 3D segmentations. As exhibited in Figure~\ref{fig:eyeballdistri}, most of the eyeball centers are located in the image. There are, however, some eyeball centers located outside of the image (Y position greater than 144). As is the case for some of the images in the LPW data set, this is due to the camera's very close proximity to the eye. The wide variation in the eyeball's radius is likewise a result of the camera's different distances.

\section{Annotation Process}
For the annotation of the landmarks and semantic segmentation in TüEyeD, we used a semi-supervised approach together with the multiple annotation maturation (MAM)~\cite{ICCVW2018FuhlW} algorithm. Unlike the original algorithm, we used CNNs~\cite{lecun2015deep,rosenblatt1958perceptron,NORM2020FUHL} instead of SVMs~\cite{scholkopf2002learning} in combination with HOG~\cite{dalal2005histograms} features. We also limited the iterations to five and used two competing models. One model consisted of a ResNet50 and was trained for landmark regression using the validation loss function of \cite{ICMV2019FuhlW}. This loss function enabled the CNN to detect whether the pupil, iris, and eyelids were present. The loss function also provided information about the accuracy of the individual landmarks. For the other model, we trained the semantic segmentation together with a U-Net~\cite{ronneberger2015u} and residual blocks~\cite{he2016deep}. For both models, we also used the batch balancing of \cite{ICMV2019FuhlW,NNVALID2020FUHL}.

Initially, we annotated 20,000 images with landmarks and converted them into semantic segmentations. Then we trained the CNNs and continuously improved them with the MAM algorithm. After five iterations, the ResNet50 landmarks were converted into semantic segmentations and compared to the U-Net results. For this step, we used the Jaccard index, i.e., $\frac{ \text{ResNet50} \cap \text{U-Net} }{ \text{ResNet50} \cup \text{U-Net} }$. If this value was less than $0.9$, applicable images were marked and new images selected from the set for manual annotation. A total of four post-annotations were completed and the process was started again from scratch.

\textbf{3D eyeball and optical vector} were annotated based on the approach presented in \cite{NNETRA2020}. However, instead of using the pupil ellipse, we used the iris ellipse since it is only partially affected by corneal refraction. Additionally, we used both approaches from \cite{NNETRA2020}, wherein one approach processes several ellipses in a neural network and the other approach calculates single vectors from single ellipses. For the second approach, we calculated the minimum intersection point of the individual vectors and the resulting radius. In regard to segmentation, we compared both approaches and accepted deviations of less than two pixels for the center and radius of the eyeball. In all other cases, we made manual correction utilizing the preceding and succeeding eyeball parameters.

\textbf{3D landmarks and segmentation} were calculated geometrically by combining the 2D landmarks and segmentations with the 3D eyeball model. As the pupil is always physically located at the center of the iris, we accounted for two different 3D segmentations and 3D landmarks. We first considered how the pupil appears in the eye image. Due to corneal refractions and steep camera angles, the pupil appears often not in the center of the iris. Accordingly, we adjusted the 3D landmarks and 3D segmentation to the iris and, more specifically, to the center of the iris.

\textbf{Eye movements} are annotated as fixations ("still" eye), saccades (fast eye movement between fixations), smooth pursuits (slow eye movement), and  blinks. Additionally, all images without an eye or with open eyes lacking valid pupil coverage were marked as errors. In the first step of our annotation, 50,000 individual images were annotated using the optical vector annotation. Then, semantic segmentation for eye movements was applied to the angular velocities of the optical vector~\cite{FCDGR2020FUHL}. On top of it, we applied the MAM approach for two iterations. Finally, the detected eye movement types were validated against biologically valid parameters~\cite{purves2001types} and manually corrected for errors.

\section{Baseline Evaluations}

\textbf{Generalisation across eye trackers.} To highlight some of the advantages that come with this large data set, in our first baseline experiment we analyzed the generalization performance for \textbf{landmark regression} and \textbf{semantic segmentation} across different eye trackers. Note that cross-eye-tracker generalization poses a key challenge for eye-tracking manufacturers for the mentioned tasks, since, as of now, changing eye-tracking devices involves the manual annotation of images generated by the new device. 

In our experiment, we apply a leave-one-out cross validation, i.e., the data from every, but one, eye tracker is used for training. Data from the omitted eye tracker serves as the validation set. Additionally, we ensure that data from all the other eye trackers is also contained in the validation set. As a final step, the mean value for the F1 score as well as the mean Jaccard index (mJI) is calculated, where each data set is equally weighted. 
As a test data set, we hold back 6,379,400 images with annotations from the eye trackers Look! and Enke GmbH. In order to evaluate over this data, 
we used the models ResNet-34~\cite{he2016deep}, ResNet-50~\cite{he2016deep}, MobilNetV2~\cite{sandler2018mobilenetv2}, and U-Net~\cite{ronneberger2015u} with residual blocks~\cite{he2016deep} and batch normalization~\cite{ioffe2015batch}. The training data was additionally augmented with 0-30\% random noise, rotations between -45-45$^\circ$, shifts of 0-20\%, 1.0-2.0 standard deviation blure, overlaying with images to simulate reflections, adding vertical and horizontal noise to pixel lines, and adding 0-10 noisy squares or ellipses with random size and orientation. As optimizer for the semantic segmentation, we used SGD~\cite{qian1999momentum} with the parameters $5*10^{-4}$ weight decay, 0.99 momentum, and 0.1 learning rate. After every sequence one thousand epochs, the learning rate was reduced by a factor of 0.1. This was performed up to a learning rate of $10^{-5}$. As a loss function for the pixel classes softmax was used. For landmark regression, Adam~\cite{kingma2014adam} was used with the parameters $5*10^{-4}$ as weight decay, 0.9 and 0.99 for the first and second momentum, respectively, and learning rate of $10^{-4}$. After every sequence of one thousand epochs, the learning rate was reduced by a factor of 0.1. This was enacted up to a learning rate of $10^{-8}$. L2 was used as the loss function.

\textbf{Evaluation environment} We used the C++-based CuDNN framework for the neural net models. The Evaluation was performed with pre-compiled executables. The hardware for the test environment involves an Intel i5-4570 CPU with 4 cores, 16 Gigabyte DDR4 memory and an NVIDIA 1050ti with 4 Gigabyte memory. 

\definecolor{Gray}{gray}{0.8}

\begin{table}
\begin{small}
	\begin{center}
		\caption{Landmark regression results in average euclidean pixel distance divided by the image resolution diagonal and multiplied with the factor $10^{2}$ for the pupil, iris, and eyelid 3D landmarks on TüEyeD. Best results in bold.}
		\label{tbl:lmtest}
		\setlength\tabcolsep{4pt}
		\begin{tabular}{cccccc}
		\\\hline
			Model & Train Data & Res. & Pupil & Iris & Eyelid  \\ \hline
			 & & ${px}$ & $\frac{px}{diag}$ & $\frac{px}{diag}$ & $\frac{px}{diag}$  \\ \hline
			 \rowcolor{Gray} & & $192 \times 144$ & 3.23 & 2.70 & 2.51  \\
			 \rowcolor{Gray} & & $384 \times 288$ & 2.92 & 2.48 & 2.19  \\
			 \rowcolor{Gray} \multirow{-3}{*}{\rotatebox{90}{Res-34}} & 
			 \multirow{-3}{*}{\rotatebox{90}{LPW}} & $768 \times 576$ & 2.45 & 2.07 & 1.83  \\ \hline
			 \rowcolor{Gray}  &   & $192 \times 144$ & 2.15 & 1.85 & 1.55  \\ 
			 \rowcolor{Gray} & & $384 \times 288$ & 1.89 & 1.63 & 1.25  \\
		     \rowcolor{Gray} \multirow{-3}{*}{\rotatebox{90}{Res-34}} & \multirow{-3}{*}{\rotatebox{90}{ALL}} & $768 \times 576$ & 1.76 & 1.46 & 1.13  \\ \hline
			\multirow{3}{*}{\rotatebox{90}{Res-50}}& \multirow{3}{*}{\rotatebox{90}{ALL}} & $192 \times 144$ & 2.02 & 1.65 & 1.34  \\
			& & $384 \times 288$ & 1.68 & 1.46 & 1.08  \\
			& & $768 \times 576$ & \textbf{1.54} & \textbf{1.34} & \textbf{1.01}  \\ \hline
			\multirow{3}{*}{\rotatebox{90}{MobV2}}& \multirow{3}{*}{\rotatebox{90}{ALL}} & $192 \times 144$ & 2.50 & 2.01 & 1.62  \\
			& & $384 \times 288$ & 2.11 & 1.73 & 1.40  \\
			& & $768 \times 576$ & 1.94 & 1.54 & 1.20  \\ \hline
		\end{tabular}
	\end{center}
	\end{small}
\end{table}

\textbf{Results on landmark regression.} Table~\ref{tbl:lmtest} shows the results of the landmark regression. For this purpose, we trained different models that determine landmarks for the pupil, iris, and eyelids together. Note, however, that the results can be further improved by using individual models for estimating the landmarks of the pupil, iris, and the eyelids. This is largely because eyelids move independently of the pupil and the iris, and the pupil is displaced from the iris due to corneal refraction. As an evaluation measure we report the mean distance of the predictions from the ground truth annotations, as pixels normalized by the diagonal (as $\frac{px}{diag}$). Table~\ref{tbl:lmtest} shows, as expected, that larger models are more effective on the described regression tasks. The same conclusion can be drawn from Table~\ref{tbl:eyeballtest}, where the results of the eyeball parameter estimation are shown. For this purpose, we trained different models, each having received five consecutive images as input. Also in this case, larger models and higher resolutions are more effective. However, in both Tables~\ref{tbl:lmtest} and~\ref{tbl:eyeballtest} we can see the clear advantage of the TüEyeD data set in comparison to smaller existing data sets, as depicted by the topmost two models (highlighted in gray) which use the same model architecture, i.e., ResNet-34, but are trained once on the LPW data set and once the full TüEyeD data set. Furthermore, the results also indicate, as expected, that cross-eye-tracker generalization on images taken in real-world settings is a challenging task, which however can be approached using TüEyeD together with more complex architectures. Thus, now the key challenge of cross-eye-tracker generalization can be easily approached without the need for creating and annotating new data, whenever a new eye-tracking device is used.

\begin{table}
\begin{small}
	\begin{center}
		\caption{Eyeball parameter and gaze vector (GV) regression results in average euclidean pixel distance divided by the image resolution diagonal and multiplied with the factor $10^{2}$ for the 3D position as well as for the radius and average angular difference for the gaze vector on TüEyeD. Each model received five consecutive images as input to estimate the eye ball parameters and the current gaze vector. Best results in bold.}
		\label{tbl:eyeballtest}
		\setlength\tabcolsep{4pt}
		\begin{tabular}{cccccc}
		\\\hline
			Model & Train Data & Res. & Eyeball & Radius & GV   \\ \hline
			 & & ${px}$ & $\frac{px}{diag}$ & $\frac{px}{diag}$ & degree   \\ \hline
			 \rowcolor{Gray} &  & $192 \times 144$ & 2.41 & 2.19 & 5.37 \\
			 \rowcolor{Gray} &  & $384 \times 288$  & 2.16 & 2.02 & 4.90  \\
			 \rowcolor{Gray} \multirow{-3}{*}{\rotatebox{90}{Res-34}} & \multirow{-3}{*}{\rotatebox{90}{LPW}} & $768 \times 576$  & 1.80 & 1.93 & 4.41  \\ \hline
			 \rowcolor{Gray} &  & $192 \times 144$ & 1.74 & 1.35 & 4.72 \\
			  \rowcolor{Gray}& & $384 \times 288$  & 1.54 & 1.22 & 3.92  \\
			  \rowcolor{Gray}\multirow{-3}{*}{\rotatebox{90}{Res-34}} & \multirow{-3}{*}{\rotatebox{90}{ALL}} & $768 \times 576$  & 1.32 & 1.03 & 3.18  \\ \hline
			\multirow{3}{*}{\rotatebox{90}{Res-50}} & \multirow{3}{*}{\rotatebox{90}{ALL}} & $192 \times 144$ & 1.54 & 1.22 & 4.15 \\
			 & & $384 \times 288$  & 1.25 & 1.06 & 3.69  \\
			 & & $768 \times 576$  & \textbf{1.05} & \textbf{0.96} & \textbf{2.97} \\ \hline
			\multirow{3}{*}{\rotatebox{90}{MobV2}} & \multirow{3}{*}{\rotatebox{90}{ALL}} & $192 \times 144$ & 1.95 & 1.72 & 5.05 \\
			 & & $384 \times 288$  & 1.73 & 1.62 & 4.19 \\
			 & & $768 \times 576$  & 1.55 & 1.25 & 3.57  \\ \hline
		\end{tabular}
	\end{center}
\end{small}
\end{table}
\begin{table}
\begin{small}
	\begin{center}
		\caption{Semantic segmentation results as mean Jaccard index (mJI) on the test set. For the models ResNet-34 (Res-34), ResNet-50 (Res-50), and MbileNetV2 (MobV2) we converted the landmarks into segments using OpenCV~\cite{opencvlibrary}. Z is the average euclidean distance divided by the image resolution diagonal and multiplied with the factor $10^{2}$ for the 3D position of the segments. Best results in bold.}
		\label{tbl:semsegtest}
		\setlength\tabcolsep{4pt}
		\begin{tabular}{cccccc}
		\\\hline
			 Model & Res. & Pupil & Iris & Sclera & Z  \\ \hline
			  & px & mJI & mJI & mJI & $\frac{px}{diag}$  \\ \hline
			 \multirow{3}{*}{\rotatebox{90}{U-Net}} & $192 \times 144$ & 0.52 & 0.58 & 0.67 & 2.73 \\
			  & $384 \times 288$ & 0.56 & 0.61 & 0.69 & 2.49 \\
			  & $768 \times 576$ & 0.60 & 0.62 & 0.70 & 1.95 \\ \hline
			  \multirow{3}{*}{\rotatebox{90}{Res-34}} & $192 \times 144$ & 0.58 & 0.62 & 0.71 & 1.90 \\
			  & $384 \times 288$ & 0.60 & 0.64 & 0.73 & 1.45 \\
			  & $768 \times 576$ & 0.63 & 0.68 & 0.76 & 1.12 \\ \hline
			  \multirow{3}{*}{\rotatebox{90}{Res-50}} & $192 \times 144$ & 0.61 & 0.65 & 0.75 & 1.71 \\
			  & $384 \times 288$ & 0.64 & 0.66 & 0.77 & 1.22 \\
			  & $768 \times 576$ & \textbf{0.65} & \textbf{0.70} & \textbf{0.78} & \textbf{0.85} \\ \hline
			  \multirow{3}{*}{\rotatebox{90}{MobV2}} & $192 \times 144$ & 0.56 & 0.60 & 0.70 & 2.07 \\
			  & $384 \times 288$ & 0.59 & 0.62 & 0.72 & 1.75 \\
			  & $768 \times 576$ & 0.61 & 0.65 & 0.74 & 1.46 \\ \hline
		\end{tabular}
	\end{center}
	\end{small}
\end{table}


\textbf{Semantic segmentation.} Table~\ref{tbl:semsegtest} shows the results for semantic segmentation. For the landmark regression models, we created the semantic segments using the OpenCV ellipse fit for the iris and the pupil, and the fillPolygon function for the eyelids. Also for this task, we conclude that despite the challenging eye images from different eye trackers and real-world scenarios, a fairly viable generalization can be achieved using TüEyeD together with larger models.

\begin{table}
\begin{small}
	\begin{center}
		\caption{Eye movement segmentation results are provided as the mean Jaccard Index (mJI) on the test set. The models ResNet-34, ResNet-50, and MobileNetV2 are used in a window-based fashion on 256 consecutive input values from the ground truth, i.e., pupil center (PC) or gaze vector (GV), and predict on 16 consecutive data points, i.e. the corresponding eye movement events: Fixations (Fix.), Saccades (Sacc.), Smooth Pursuits (Sm.Purs.), Errors and Blinks. Best results are highlighted in bold.}
		\label{tbl:movenetsclassify}
		\setlength\tabcolsep{4pt}
		\begin{tabular}{ccccccc}
		\\\hline
			Input & Model & Fix. & Sacc. & Sm.Purs. & Error & Blink  \\ \hline
			 \multirow{3}{*}{\rotatebox{90}{PC}} & ResNet-34 & 0.81 & 0.73 & 0.83 & 0.92 & 0.81 \\
			 & ResNet-50 & \textbf{0.86} & \textbf{0.75} & \textbf{0.87} & \textbf{0.95} & \textbf{0.83} \\
			 & MobileNetV2 & 0.78 & 0.70 & 0.81 & 0.89 & 0.74 \\ \hline
			 \multirow{3}{*}{\rotatebox{90}{GV}} & ResNet-34 & 0.92 & 0.87 & 0.91 & \textbf{0.98} & 0.90 \\
			 & ResNet-50 & \textbf{0.94} & \textbf{0.89} & \textbf{0.93} & \textbf{0.98} & \textbf{0.91} \\
			 & MobileNetV2 & 0.85 & 0.82 & 0.89 & 0.97 & 0.88 \\ \hline
		\end{tabular}
	\end{center}
	\end{small}
\end{table}
\textbf{Recognition of eye movement types.} Table~\ref{tbl:movenetsclassify} presents the results on eye movement recognition. The models had to predict the eye movement type in addition to the errors and blinks. For this purpose, they received the ground truth of the pupil in the upper part of the evaluation and the gaze vector in the second part of the evaluation. All models were applied in a window-based fashion and received 256 data points (raw eye-tacking data points) to classify 16 data points. These 16 data points to be predicted were exactly in the middle of the 256 data points. For each data point, we also appended the time in milliseconds (ms) from the previous data point. As it can be seen, the gaze vector (GV) is much more effective for eye-movement classification because it compensates for shifts of the eye tracker. Due to the difficulty of computing a robust signal of the gaze vector, the pupil center is still taken in conventional systems. Also in this case, TüEyeD can be used to achieve the generalization across different eye trackers and different real-world settings.

\section{Conclusion}
In this work, we presented TüEyeD, a rich and coherent data set of over 20 Million eye images along with their 2D and 3D annotations and other annotations including eye movement types, semantic segmentations, landmarks, elliptical parameters for the iris, the pupil, and the eyelid as well as eyeball parameters for shift invariant gaze estimation. Generated by a total of seven different eye trackers with different sampling rates and under challenging real-world conditions, TüEyeD is the most comprehensive and realistic data set of semantically annotated eye images to date. This data set should not only be seen as a new foundational resource in the field of computer vision and eye movement research. We are convinced that TüEyeD will also have a profound impact on other fields and communities, ranging from cognitive sciences to AR and VR applications as well as novel visualization techniques~\cite{TCKWGJRWE2015,ROIGA2018,ASAOIB2015}. At the very least, it will unquestionably contribute to the application of eye-movement and gaze estimation techniques in challenging practical use cases.

{\small
\bibliographystyle{ieeefullname}
\bibliography{egbib}

\begin{thebibliography}{10}\itemsep=-1pt

\bibitem{adjouadi2004remote}
Malek Adjouadi, Anaelis Sesin, Melvin Ayala, and Mercedes Cabrerizo.
\newblock Remote eye gaze tracking system as a computer interface for persons
  with severe motor disability.
\newblock In {\em International conference on computers for handicapped
  persons}, pages 761--769. Springer, 2004.

\bibitem{Bahmani2016}
H. Bahmani, W. Fuhl, E. Gutierrez, G. Kasneci, E. Kasneci, and S. Wahl.
\newblock Feature-based attentional influences on the accommodation response.
\newblock In {\em Vision Sciences Society Annual Meeting Abstract}, 2016.

\bibitem{bergmuller2014impact}
Thomas Bergm{\"u}ller, Luca Debiasi, Andreas Uhl, and Zhenan Sun.
\newblock Impact of sensor ageing on iris recognition.
\newblock In {\em IEEE International Joint Conference on Biometrics}, pages
  1--8. IEEE, 2014.

\bibitem{boles1998security}
Wageeh~W Boles.
\newblock A security system based on human iris identification using wavelet
  transform.
\newblock {\em Engineering Applications of Artificial Intelligence},
  11(1):77--85, 1998.

\bibitem{opencvlibrary}
G. Bradski.
\newblock {The OpenCV Library}.
\newblock {\em Dr. Dobb's Journal of Software Tools}, 2000.

\bibitem{breiman2001random}
Leo Breiman.
\newblock Random forests.
\newblock {\em Machine learning}, 45(1):5--32, 2001.

\bibitem{chen2016xgboost}
Tianqi Chen and Carlos Guestrin.
\newblock Xgboost: A scalable tree boosting system.
\newblock In {\em Proceedings of the 22nd acm sigkdd international conference
  on knowledge discovery and data mining}, pages 785--794, 2016.

\bibitem{dalal2005histograms}
Navneet Dalal and Bill Triggs.
\newblock Histograms of oriented gradients for human detection.
\newblock In {\em 2005 IEEE computer society conference on computer vision and
  pattern recognition (CVPR'05)}, volume~1, pages 886--893. IEEE, 2005.

\bibitem{dalton2005gaze}
Kim~M Dalton, Brendon~M Nacewicz, Tom Johnstone, Hillary~S Schaefer, Morton~Ann
  Gernsbacher, Hill~H Goldsmith, Andrew~L Alexander, and Richard~J Davidson.
\newblock Gaze fixation and the neural circuitry of face processing in autism.
\newblock {\em Nature neuroscience}, 8(4):519--526, 2005.

\bibitem{das2019sclera}
Abhijit Das, Umapada Pal, Michael Blumenstein, Caiyong Wang, Yong He, Yuhao
  Zhu, and Zhenan Sun.
\newblock Sclera segmentation benchmarking competition in cross-resolution
  environment.
\newblock In {\em 2019 International Conference on Biometrics (ICB)}, pages
  1--7. IEEE, 2019.

\bibitem{das2016ssrbc}
Abhijit Das, Umapada Pal, Miguel~A Ferrer, and Michael Blumenstein.
\newblock Ssrbc 2016: Sclera segmentation and recognition benchmarking
  competition.
\newblock In {\em 2016 International Conference on Biometrics (ICB)}, pages
  1--6. IEEE, 2016.

\bibitem{das2017sserbc}
Abhijit Das, Umapada Pal, Miguel~A Ferrer, Michael Blumenstein, Dejan
  {\v{S}}tepec, Peter Rot, {\v{Z}}iga Emer{\v{s}}i{\v{c}}, Peter Peer, Vitomir
  {\v{S}}truc, SV~Aruna Kumar, et~al.
\newblock Sserbc 2017: Sclera segmentation and eye recognition benchmarking
  competition.
\newblock In {\em 2017 IEEE International Joint Conference on Biometrics
  (IJCB)}, pages 742--747. IEEE, 2017.

\bibitem{david2018dataset}
Erwan~J David, Jes{\'u}s Guti{\'e}rrez, Antoine Coutrot, Matthieu~Perreira
  Da~Silva, and Patrick~Le Callet.
\newblock A dataset of head and eye movements for 360 videos.
\newblock In {\em Proceedings of the 9th ACM Multimedia Systems Conference},
  pages 432--437, 2018.

\bibitem{dinges2005pilot}
David~F Dinges, Greg Maislin, Rebecca~M Brewster, Gerald~P Krueger, and
  Robert~J Carroll.
\newblock Pilot test of fatigue management technologies.
\newblock {\em Transportation research record}, 1922(1):175--182, 2005.

\bibitem{dong2005fatigue}
Wenhui Dong and Xiaojuan Wu.
\newblock Fatigue detection based on the distance of eyelid.
\newblock In {\em Proceedings of 2005 IEEE International Workshop on VLSI
  Design and Video Technology, 2005.}, pages 365--368. IEEE, 2005.

\bibitem{duchowski2018index}
Andrew~T Duchowski, Krzysztof Krejtz, Izabela Krejtz, Cezary Biele, Anna
  Niedzielska, Peter Kiefer, Martin Raubal, and Ioannis Giannopoulos.
\newblock The index of pupillary activity: Measuring cognitive load
  vis-{\`a}-vis task difficulty with pupil oscillation.
\newblock In {\em Proceedings of the 2018 CHI Conference on Human Factors in
  Computing Systems}, pages 1--13, 2018.

\bibitem{0320170}
S Eivazi, W Fuhl, and E Kasneci.
\newblock Towards intelligent surgical microscopes: Surgeons gaze and
  instrument tracking.
\newblock In {\em Proceedings of the 22st International Conference on
  Intelligent User Interfaces, IUI}, 2017.

\bibitem{ACTNEURO2017}
S. Eivazi, A. Hafez, W. Fuhl, H. Afkari, E. Kasneci, M. Lehecka, and R.
  Bednarik.
\newblock Optimal eye movement strategies: a comparison of neurosurgeons gaze
  patterns when using a surgical microscope.
\newblock {\em Acta Neurochirurgica}, 2017.

\bibitem{032017}
Shahram Eivazi, Michael Slupina, Wolfgang Fuhl, Hoorieh Afkari, Ahmad Hafez,
  and Enkelejda Kasneci.
\newblock Towards automatic skill evaluation in microsurgery.
\newblock In {\em Proceedings of the 22st International Conference on
  Intelligent User Interfaces, IUI 2017}. ACM, 03 2017.

\bibitem{freund1996experiments}
Yoav Freund, Robert~E Schapire, et~al.
\newblock Experiments with a new boosting algorithm.
\newblock In {\em icml}, volume~96, pages 148--156. Citeseer, 1996.

\bibitem{friedman2002stochastic}
Jerome~H Friedman.
\newblock Stochastic gradient boosting.
\newblock {\em Computational statistics \& data analysis}, 38(4):367--378,
  2002.

\bibitem{WF042019}
W. Fuhl.
\newblock {\em Image-based extraction of eye features for robust eye tracking}.
\newblock PhD thesis, University of Tübingen, 04 2019.

\bibitem{UMUAI2020FUHL}
Wolfgang Fuhl.
\newblock From perception to action using observed actions to learn gestures.
\newblock {\em User Modeling and User-Adapted Interaction}, pages 1--18, 08
  2020.

\bibitem{C2019}
Wolfgang Fuhl, Efe Bozkir, Benedikt Hosp, Nora Castner, David Geisler, Thiago
  C., and Enkelejda Kasneci.
\newblock Encodji: Encoding gaze data into emoji space for an amusing scanpath
  classification approach ;).
\newblock In {\em Eye Tracking Research and Applications}, 2019.

\bibitem{RLDIFFPRIV2020FUHL}
Wolfgang Fuhl, Efe Bozkir, and Enkelejda Kasneci.
\newblock Reinforcement learning for the privacy preservation and manipulation
  of eye tracking data.
\newblock {\em arXiv preprint arXiv:2002.06806}, 08 2020.

\bibitem{ICMIW2019FuhlW1}
W. Fuhl, N. Castner, and E. Kasneci.
\newblock Histogram of oriented velocities for eye movement detection.
\newblock In {\em International Conference on Multimodal Interaction Workshops,
  ICMIW}, 2018.

\bibitem{ICMIW2019FuhlW2}
W. Fuhl, N. Castner, and E. Kasneci.
\newblock Rule based learning for eye movement type detection.
\newblock In {\em International Conference on Multimodal Interaction Workshops,
  ICMIW}, 2018.

\bibitem{FFAO2019}
W. Fuhl, N. Castner, T.~C. Kübler, A. Lotz, W. Rosenstiel, and E. Kasneci.
\newblock Ferns for area of interest free scanpath classification.
\newblock In {\em Proceedings of the 2019 ACM Symposium on Eye Tracking
  Research \& Applications (ETRA)}, 06 2019.

\bibitem{ICCVW2018FuhlW}
W. Fuhl, N. Castner, L. Zhuang, M. Holzer, W. Rosenstiel, and E. Kasneci.
\newblock Mam: Transfer learning for fully automatic video annotation and
  specialized detector creation.
\newblock In {\em International Conference on Computer Vision Workshops,
  ICCVW}, 2018.

\bibitem{ETRA2018FuhlW}
W. Fuhl, S. Eivazi, B. Hosp, A. Eivazi, W. Rosenstiel, and E. Kasneci.
\newblock Bore: Boosted-oriented edge optimization for robust, real time remote
  pupil center detection.
\newblock In {\em Eye Tracking Research and Applications, ETRA}, 2018.

\bibitem{NNETRA2020}
W. Fuhl, H. Gao, and E. Kasneci.
\newblock Neural networks for optical vector and eye ball parameter estimation.
\newblock In {\em ACM Symposium on Eye Tracking Research \& Applications, ETRA
  2020}. ACM, 01 2020.

\bibitem{VECETRA2020}
W. Fuhl, H. Gao, and E. Kasneci.
\newblock Tiny convolution, decision tree, and binary neuronal networks for
  robust and real time pupil outline estimation.
\newblock In {\em ACM Symposium on Eye Tracking Research \& Applications, ETRA
  2020}. ACM, 01 2020.

\bibitem{ICCVW2019FuhlW}
W. Fuhl, D. Geisler, W. Rosenstiel, and E. Kasneci.
\newblock The applicability of cycle gans for pupil and eyelid segmentation,
  data generation and image refinement.
\newblock In {\em International Conference on Computer Vision Workshops,
  ICCVW}, 11 2019.

\bibitem{WDTTWE062018}
W. Fuhl, D. Geisler, T. Santini, T. Appel, W. Rosenstiel, and E. Kasneci.
\newblock Cbf:circular binary features for robust and real-time pupil center
  detection.
\newblock In {\em ACM Symposium on Eye Tracking Research \& Applications}, 06
  2018.

\bibitem{WDTE092016}
W. Fuhl, D. Geisler, T. Santini, and E. Kasneci.
\newblock Evaluation of state-of-the-art pupil detection algorithms on remote
  eye images.
\newblock In {\em ACM International Joint Conference on Pervasive and
  Ubiquitous Computing: Adjunct publication -- PETMEI 2016}, 09 2016.

\bibitem{EPIC2018FuhlW}
W. Fuhl and E. Kasneci.
\newblock Eye movement velocity and gaze data generator for evaluation,
  robustness testing and assess of eye tracking software and visualization
  tools.
\newblock In {\em Poster at Egocentric Perception, Interaction and Computing,
  EPIC}, 2018.

\bibitem{ICMV2019FuhlW}
W. Fuhl and E. Kasneci.
\newblock Learning to validate the quality of detected landmarks.
\newblock In {\em International Conference on Machine Vision, ICMV}, 11 2019.

\bibitem{NNPOOL2020FUHL}
Wolfgang Fuhl and Enkelejda Kasneci.
\newblock Multi layer neural networks as replacement for pooling operations.
\newblock {\em arXiv preprint arXiv:2006.06969}, 08 2020.

\bibitem{RINGRAD2020FUHL}
Wolfgang Fuhl and Enkelejda Kasneci.
\newblock Rotated ring, radial and depth wise separable radial convolutions.
\newblock {\em arXiv}, 08 2020.

\bibitem{NORM2020FUHL}
Wolfgang Fuhl and Enkelejda Kasneci.
\newblock Weight and gradient centralization in deep neural networks.
\newblock {\em arXiv}, 08 2020.

\bibitem{AAAIFuhlW}
W. Fuhl, G. Kasneci, W. Rosenstiel, and E. Kasneci.
\newblock Training decision trees as replacement for convolution layers.
\newblock In {\em Conference on Artificial Intelligence, AAAI}, 02 2020.

\bibitem{AGAS2018}
W. Fuhl, T. Kübler, T. Santini, and E. Kasneci.
\newblock Automatic generation of saliency-based areas of interest.
\newblock In {\em Symposium on Vision, Modeling and Visualization (VMV)}, 09
  2018.

\bibitem{ROIGA2018}
W. Fuhl, T.~C. Kübler, H. Brinkmann, R. Rosenberg, W. Rosenstiel, and E.
  Kasneci.
\newblock Region of interest generation algorithms for eye tracking data.
\newblock In {\em Third Workshop on Eye Tracking and Visualization (ETVIS), in
  conjunction with ACM ETRA}, 06 2018.

\bibitem{WTCDOWE052017}
W. Fuhl, T.~C. Kübler, D. Hospach, O. Bringmann, W. Rosenstiel, and E.
  Kasneci.
\newblock Ways of improving the precision of eye tracking data: Controlling the
  influence of dirt and dust on pupil detection.
\newblock {\em Journal of Eye Movement Research}, 10(3), 05 2017.

\bibitem{ASAOIB2015}
W. Fuhl, T.~C. Kübler, K. Sippel, W. Rosenstiel, and E. Kasneci.
\newblock Arbitrarily shaped areas of interest based on gaze density gradient.
\newblock In {\em European Conference on Eye Movements, ECEM 2015}, 08 2015.

\bibitem{WTCKWE092015}
W. Fuhl, T.~C. Kübler, K. Sippel, W. Rosenstiel, and E. Kasneci.
\newblock Excuse: Robust pupil detection in real-world scenarios.
\newblock In {\em 16th International Conference on Computer Analysis of Images
  and Patterns (CAIP 2015)}, 09 2015.

\bibitem{FCDGR2020FUHL}
Wolfgang Fuhl, Yao Rong, and Kasneci Enkelejda.
\newblock Fully convolutional neural networks for raw eye tracking data
  segmentation, generation, and reconstruction.
\newblock In {\em Proceedings of the International Conference on Pattern
  Recognition}, pages 0--0, 2020.

\bibitem{NNVALID2020FUHL}
Wolfgang Fuhl, Yao Rong, Thomas Motz, Michael Scheidt, Andreas Hartel, Andreas
  Koch, and Enkelejda Kasneci.
\newblock Explainable online validation of machine learning models for
  practical applications.
\newblock {\em arXiv}, 08 2020.

\bibitem{CAIP2019FuhlW}
W. Fuhl, W. Rosenstiel, and E. Kasneci.
\newblock 500,000 images closer to eyelid and pupil segmentation.
\newblock In {\em Computer Analysis of Images and Patterns, CAIP}, 11 2019.

\bibitem{WTDTE022017}
W. Fuhl, T. Santini, D. Geisler, T.~C. Kübler, and E. Kasneci.
\newblock Eyelad: Remote eye tracking image labeling tool.
\newblock In {\em 12th Joint Conference on Computer Vision, Imaging and
  Computer Graphics Theory and Applications (VISIGRAPP 2017)}, 02 2017.

\bibitem{WTDTWE092016}
W. Fuhl, T. Santini, D. Geisler, T.~C. Kübler, W. Rosenstiel, and E. Kasneci.
\newblock Eyes wide open? eyelid location and eye aperture estimation for
  pervasive eye tracking in real-world scenarios.
\newblock In {\em ACM International Joint Conference on Pervasive and
  Ubiquitous Computing: Adjunct publication -- PETMEI 2016}, 09 2016.

\bibitem{WTE032017}
W. Fuhl, T. Santini, and E. Kasneci.
\newblock Fast and robust eyelid outline and aperture detection in real-world
  scenarios.
\newblock In {\em IEEE Winter Conference on Applications of Computer Vision
  (WACV 2017)}, 03 2017.

\bibitem{CORR2017FuhlW1}
W. Fuhl, T. Santini, and E. Kasneci.
\newblock Fast camera focus estimation for gaze-based focus control.
\newblock In {\em CoRR}, 2017.

\bibitem{CORR2016FuhlW}
W. Fuhl, T. Santini, G. Kasneci, and E. Kasneci.
\newblock Pupilnet: Convolutional neural networks for robust pupil detection.
\newblock In {\em CoRR}, 2016.

\bibitem{CORR2017FuhlW2}
W. Fuhl, T. Santini, G. Kasneci, and E. Kasneci.
\newblock Pupilnet v2.0: Convolutional neural networks for robust pupil
  detection.
\newblock In {\em CoRR}, 2017.

\bibitem{WTTE032016}
W. Fuhl, T. Santini, T.~C. Kübler, and E. Kasneci.
\newblock Else: Ellipse selection for robust pupil detection in real-world
  environments.
\newblock In {\em Proceedings of the Ninth Biennial ACM Symposium on Eye
  Tracking Research \& Applications (ETRA)}, pages 123--130, 03 2016.

\bibitem{fuhl2018simarxiv}
W. Fuhl, T. Santini, T. Kuebler, N. Castner, W. Rosenstiel, and E. Kasneci.
\newblock Eye movement simulation and detector creation to reduce laborious
  parameter adjustments.
\newblock {\em arXiv preprint arXiv:1804.00970}, 2018.

\bibitem{WTCDAHKSE122016}
W. Fuhl, T. Santini, C. Reichert, D. Claus, A. Herkommer, H. Bahmani, K. Rifai,
  S. Wahl, and E. Kasneci.
\newblock Non-intrusive practitioner pupil detection for unmodified microscope
  oculars.
\newblock {\em Elsevier Computers in Biology and Medicine}, 79:36--44, 12 2016.

\bibitem{fuhl2016pupil}
Wolfgang Fuhl, Marc Tonsen, Andreas Bulling, and Enkelejda Kasneci.
\newblock Pupil detection for head-mounted eye tracking in the wild: an
  evaluation of the state of the art.
\newblock {\em Machine Vision and Applications}, 27(8):1275--1288, 2016.

\bibitem{062016}
Wolfgang Fuhl, Marc Tonsen, Andreas Bulling, and Enkelejda Kasneci.
\newblock Pupil detection for head-mounted eye tracking in the wild: An
  evaluation of the state of the art.
\newblock In {\em Machine Vision and Applications}, pages 1--14, 06 2016.

\bibitem{garbin2020dataset}
Stephan~Joachim Garbin, Oleg Komogortsev, Robert Cavin, Gregory Hughes, Yiru
  Shen, Immo Schuetz, and Sachin~S Talathi.
\newblock Dataset for eye tracking on a virtual reality platform.
\newblock In {\em ACM Symposium on Eye Tracking Research and Applications},
  pages 1--10, 2020.

\bibitem{DWTE022017}
D. Geisler, W. Fuhl, T. Santini, and E. Kasneci.
\newblock Saliency sandbox: Bottom-up saliency framework.
\newblock In {\em 12th Joint Conference on Computer Vision, Imaging and
  Computer Graphics Theory and Applications (VISIGRAPP 2017)}, 02 2017.

\bibitem{he2016deep}
Kaiming He, Xiangyu Zhang, Shaoqing Ren, and Jian Sun.
\newblock Deep residual learning for image recognition.
\newblock In {\em Proceedings of the IEEE conference on computer vision and
  pattern recognition}, pages 770--778, 2016.

\bibitem{hooker2005you}
Christine Hooker and Sohee Park.
\newblock You must be looking at me: The nature of gaze perception in
  schizophrenia patients.
\newblock {\em Cognitive neuropsychiatry}, 10(5):327--345, 2005.

\bibitem{hutchinson1989human}
Thomas~E Hutchinson, K~Preston White, Worthy~N Martin, Kelly~C Reichert, and
  Lisa~A Frey.
\newblock Human-computer interaction using eye-gaze input.
\newblock {\em IEEE Transactions on systems, man, and cybernetics},
  19(6):1527--1534, 1989.

\bibitem{ioffe2015batch}
Sergey Ioffe and Christian Szegedy.
\newblock Batch normalization: Accelerating deep network training by reducing
  internal covariate shift.
\newblock {\em arXiv preprint arXiv:1502.03167}, 2015.

\bibitem{ishiyama2015usefulness}
Yukako Ishiyama, Hiroshi Murata, and Ryo Asaoka.
\newblock The usefulness of gaze tracking as an index of visual field
  reliability in glaucoma patients.
\newblock {\em Investigative ophthalmology \& visual science},
  56(11):6233--6236, 2015.

\bibitem{kasneci2014driving}
Enkelejda Kasneci, Katrin Sippel, Kathrin Aehling, Martin Heister, Wolfgang
  Rosenstiel, Ulrich Schiefer, and Elena Papageorgiou.
\newblock Driving with binocular visual field loss? a study on a supervised
  on-road parcours with simultaneous eye and head tracking.
\newblock {\em PloS one}, 9(2):e87470, 2014.

\bibitem{TCKWGJRWE2015}
T.~C. Kübler, K. Sippel, W. Fuhl, G. Schievelbein, J. Aufreiter, R. Rosenberg,
  W. Rosenstiel, and E. Kasneci.
\newblock {\em Analysis of eye movements with Eyetrace}, volume 574.
\newblock Biomedical Engineering Systems and Technologies. Communications in
  Computer and Information Science (CCIS). Springer International Publishing,
  2015.

\bibitem{kim2019nvgaze}
Joohwan Kim, Michael Stengel, Alexander Majercik, Shalini De~Mello, David Dunn,
  Samuli Laine, Morgan McGuire, and David Luebke.
\newblock Nvgaze: An anatomically-informed dataset for low-latency, near-eye
  gaze estimation.
\newblock In {\em Proceedings of the 2019 CHI Conference on Human Factors in
  Computing Systems}, pages 1--12, 2019.

\bibitem{kingma2014adam}
Diederik~P Kingma and Jimmy Ba.
\newblock Adam: A method for stochastic optimization.
\newblock {\em arXiv preprint arXiv:1412.6980}, 2014.

\bibitem{kothari2020gaze}
Rakshit Kothari, Zhizhuo Yang, Christopher Kanan, Reynold Bailey, Jeff~B Pelz,
  and Gabriel~J Diaz.
\newblock Gaze-in-wild: A dataset for studying eye and head coordination in
  everyday activities.
\newblock {\em Scientific reports}, 10(1):1--18, 2020.

\bibitem{lecun2015deep}
Yann LeCun, Yoshua Bengio, and Geoffrey Hinton.
\newblock Deep learning.
\newblock {\em nature}, 521(7553):436--444, 2015.

\bibitem{luckiesh1937eyelid}
Matthew Luckiesh and Frank~K Moss.
\newblock The eyelid reflex as a criterion of ocular fatigue.
\newblock {\em Journal of experimental Psychology}, 20(6):589, 1937.

\bibitem{mcmurrough2012eye}
Christopher~D McMurrough, Vangelis Metsis, Jonathan Rich, and Fillia Makedon.
\newblock An eye tracking dataset for point of gaze detection.
\newblock In {\em Proceedings of the Symposium on Eye Tracking Research and
  Applications}, pages 305--308, 2012.

\bibitem{patney2016towards}
Anjul Patney, Marco Salvi, Joohwan Kim, Anton Kaplanyan, Chris Wyman, Nir
  Benty, David Luebke, and Aaron Lefohn.
\newblock Towards foveated rendering for gaze-tracked virtual reality.
\newblock {\em ACM Transactions on Graphics (TOG)}, 35(6):179, 2016.

\bibitem{phillips2007comments}
P~Jonathon Phillips, Kevin~W Bowyer, and Patrick~J Flynn.
\newblock Comments on the casia version 1.0 iris data set.
\newblock {\em IEEE Transactions on Pattern Analysis and Machine Intelligence},
  29(10):1869--1870, 2007.

\bibitem{proencca2005ubiris}
Hugo Proen{\c{c}}a and Lu{\'\i}s~A Alexandre.
\newblock Ubiris: A noisy iris image database.
\newblock In {\em International Conference on Image Analysis and Processing},
  pages 970--977. Springer, 2005.

\bibitem{proenca2009ubiris}
Hugo Proenca, Silvio Filipe, Ricardo Santos, Joao Oliveira, and Luis~A
  Alexandre.
\newblock The ubiris. v2: A database of visible wavelength iris images captured
  on-the-move and at-a-distance.
\newblock {\em IEEE Transactions on Pattern Analysis and Machine Intelligence},
  32(8):1529--1535, 2009.

\bibitem{purves2001types}
Dale Purves, George~J Augustine, David Fitzpatrick, Lawrence~C Katz,
  Anthony-Samuel LaMantia, James~O McNamara, S~Mark Williams, et~al.
\newblock Types of eye movements and their functions.
\newblock {\em Neuroscience}, pages 361--390, 2001.

\bibitem{qian1999momentum}
Ning Qian.
\newblock On the momentum term in gradient descent learning algorithms.
\newblock {\em Neural networks}, 12(1):145--151, 1999.

\bibitem{rai2017dataset}
Yashas Rai, Jes{\'u}s Guti{\'e}rrez, and Patrick Le~Callet.
\newblock A dataset of head and eye movements for 360 degree images.
\newblock In {\em Proceedings of the 8th ACM on Multimedia Systems Conference},
  pages 205--210, 2017.

\bibitem{ronneberger2015u}
Olaf Ronneberger, Philipp Fischer, and Thomas Brox.
\newblock U-net: Convolutional networks for biomedical image segmentation.
\newblock In {\em International Conference on Medical image computing and
  computer-assisted intervention}, pages 234--241. Springer, 2015.

\bibitem{rosenblatt1958perceptron}
Frank Rosenblatt.
\newblock The perceptron: a probabilistic model for information storage and
  organization in the brain.
\newblock {\em Psychological review}, 65(6):386, 1958.

\bibitem{sandler2018mobilenetv2}
Mark Sandler, Andrew Howard, Menglong Zhu, Andrey Zhmoginov, and Liang-Chieh
  Chen.
\newblock Mobilenetv2: Inverted residuals and linear bottlenecks.
\newblock In {\em Proceedings of the IEEE conference on computer vision and
  pattern recognition}, pages 4510--4520, 2018.

\bibitem{santini2016bayesian}
Thiago Santini, Wolfgang Fuhl, Thomas K{\"u}bler, and Enkelejda Kasneci.
\newblock Bayesian identification of fixations, saccades, and smooth pursuits.
\newblock In {\em Proceedings of the Ninth Biennial ACM Symposium on Eye
  Tracking Research \& Applications}, pages 163--170, 2016.

\bibitem{scholkopf2002learning}
Bernhard Sch{\"o}lkopf, Alexander~J Smola, Francis Bach, et~al.
\newblock {\em Learning with kernels: support vector machines, regularization,
  optimization, and beyond}.
\newblock MIT press, 2002.

\bibitem{sturm2011mutual}
Virginia~E Sturm, Megan~E McCarthy, Ira Yun, Anita Madan, Joyce~W Yuan, Sarah~R
  Holley, Elizabeth~A Ascher, Adam~L Boxer, Bruce~L Miller, and Robert~W
  Levenson.
\newblock Mutual gaze in alzheimer's disease, frontotemporal and semantic
  dementia couples.
\newblock {\em Social Cognitive and Affective Neuroscience}, 6(3):359--367,
  2011.

\bibitem{swirski2012robust}
Lech {\'S}wirski, Andreas Bulling, and Neil Dodgson.
\newblock Robust real-time pupil tracking in highly off-axis images.
\newblock In {\em Proceedings of the Symposium on Eye Tracking Research and
  Applications}, pages 173--176, 2012.

\bibitem{swirski2013fully}
Lech Swirski and Neil Dodgson.
\newblock A fully-automatic, temporal approach to single camera, glint-free 3d
  eye model fitting.
\newblock {\em Proc. PETMEI}, pages 1--11, 2013.

\bibitem{swirski2014rendering}
Lech {\'S}wirski and Neil Dodgson.
\newblock Rendering synthetic ground truth images for eye tracker evaluation.
\newblock In {\em Proceedings of the Symposium on Eye Tracking Research and
  Applications}, pages 219--222, 2014.

\bibitem{tan2010efficient}
Tieniu Tan, Zhaofeng He, and Zhenan Sun.
\newblock Efficient and robust segmentation of noisy iris images for
  non-cooperative iris recognition.
\newblock {\em Image and vision computing}, 28(2):223--230, 2010.

\bibitem{tonsen2016labelled}
Marc Tonsen, Xucong Zhang, Yusuke Sugano, and Andreas Bulling.
\newblock Labelled pupils in the wild: a dataset for studying pupil detection
  in unconstrained environments.
\newblock In {\em Proceedings of the Ninth Biennial ACM Symposium on Eye
  Tracking Research \& Applications}, pages 139--142, 2016.

\bibitem{wild2015impact}
Peter Wild, James Ferryman, and Andreas Uhl.
\newblock Impact of (segmentation) quality on long vs. short-timespan
  assessments in iris recognition performance.
\newblock {\em IET Biometrics}, 4(4):227--235, 2015.

\bibitem{wood2015rendering}
Erroll Wood, Tadas Baltrusaitis, Xucong Zhang, Yusuke Sugano, Peter Robinson,
  and Andreas Bulling.
\newblock Rendering of eyes for eye-shape registration and gaze estimation.
\newblock In {\em Proceedings of the IEEE International Conference on Computer
  Vision}, pages 3756--3764, 2015.

\end{thebibliography}
}

\end{document}